# Unraveling the Reaction Mechanisms in a Chemically Amplified EUV Photoresist from a Combined Theoretical and Experimental Approach

Laura Galleni[1,2], Dhirendra P. Singh[1], Thierry Conard[1], Geoffrey Pourtois[1], Paul van der Heide[1], John Petersen[1], Kevin M. Dorney[1], Michiel J. van Setten[1]

[1]Imec, Kapeldreef 75, 3001 Leuven, Belgium
[2]Department of Chemistry, KU Leuven, Celestijnenlaan 200F, 3001 Leuven, Belgium

## ABSTRACT

Extreme ultraviolet (EUV) lithography has revolutionized high-volume manufacturing of nanoscale components, enabling the production of smaller, denser, and more energy efficient integrated circuit devices. Yet, the use of EUV light results in ionization driven chemistry within the imaging materials of lithography, the photoresists. The complex interplay of ionization, generation of primary and secondary electrons, and the subsequent chemical mechanisms leading to image formation in photoresists has been notoriously difficult to study. In this work, we deploy photoemission spectroscopy with a 92 eV EUV light source combined with first-principles simulations to unravel the chemical changes occurring during exposure in a model chemically amplified photoresist. The results reveal a surprising chemical reaction pathway, namely the EUV-induced breakdown of the photoacid generator (PAG), which is a critical component in the EUV mechanism. This previously unobserved reaction mechanism manifests as changes in intensity of the valence band peaks of the EUV photoemission spectrum, which are linked to degradation of the PAG via an advanced atomistic simulation framework. Our combined experimental and theoretical approach shows that EUV photoemission can simultaneously resolve chemical dynamics and the production of primary and secondary electrons, giving unique insights into the chemical transformation of photoresist materials. Our results pave the way for utilizing accessible, table-top EUV spectroscopy systems for observing EUV photoresist chemical dynamics, with the potential for time-resolved measurements of photoemission processes in the future.

## 1. INTRODUCTION

The successful adoption of extreme ultraviolet (EUV) lithography at the end of the last decade has pushed the continuation of Moore's law to advanced technology nodes[1–6]. EUV lithography employs EUV light sources to print nanometer-scale circuit layouts onto a photosensitive material, called photoresist. The EUV light sources used in EUV lithography typically have wavelength of 13.5 nm, corresponding to photon energy of about 92 eV. The short wavelength of EUV light compared to deep-UV (DUV) light, enables an approximately 14-fold increase of the single-print resolution (at equivalent numerical aperture and process factors), paving the way for the production of smaller, denser, and more efficient integrated circuit devices.[7] However, the introduction of EUV light sources has in turn resulted in a far more complicated exposure mechanism in photoresists, due to the higher photon energy of EUV photons compared to DUV photons. In general, when a photoresist is exposed to a light source in photolithography, the absorption of photons leads to a series of reactions that eventually modify the solubility of the resist in a developer solvent. The change in solubility is the desired outcome of the exposure step, as it allows pattern transfer by inducing contrast between the exposed and unexposed areas of the resist.

In DUV lithography, the mechanisms leading to the solubility switch are well understood and are triggered by the excitation of molecules upon photon absorption.[8] However, compared to DUV lithography, the exposure mechanism in photoresists in EUV lithography is far less understood.[8–10] In EUV exposure, the absorption of a 92 eV photon by valence levels leads to a photoemission process that generates primary electrons with high kinetic energy (~55–85 eV). These primary electrons can trigger further ionization reactions in the resist, which result in the generation of secondary electrons with lower kinetic energy (< 50 eV).[11–14] Both primary and secondary electrons can then further interact with the resist components, leading to additional reactions, such as ionization, excitation, or (dissociative) electron attachment followed by ion or neutral dissociation. These reactions eventually lead to the desired solubility switch of the photoresist. However, unstable radical byproducts can be generated as well, which can cause subsequent chemical processes beyond the desired solubility switch reaction. The complicated nature of these intertwined processes has hindered a complete understanding of the EUV exposure mechanism in photoresists, despite extensive work in this area in the past years.[8, 15–27]

Among all commercially available photoresist materials, the so-called chemically amplified resist (CAR) systems are the current workhorse of EUV lithography. CARs are composed of a polymer base containing varying amounts of photoacid generator (PAG) ion pairs and quencher molecules. Due to their commercial success, much work has been devoted to understanding the chemical transformation of CARs upon EUV exposure.[28–30] Despite the complex chemical pathways, the desired outcome in a CAR material is typically a thermally activated acid-catalyzed deprotection reaction in which a protecting group on the polymer chain is removed, thus causing a change in solubility. The acid that catalyzes the deprotection reaction is formed through the protonation of $PAG^-$ anions. This mechanism of acid formation from the PAG is believed to be started by either low energy electron attachment or electron-induced excitation of the $PAG^+$ cation followed by proton transfer to the $PAG^-$ anion.[8, 16, 31] Due to the relatively similar EUV absorption cross section of components of a CAR material, it is assumed that the primary and secondary electrons initiating this reaction can be generated from nearly any component in the resist matrix. Furthermore, direct deprotection from EUV ionization can also occur, as revealed by EUV-induced outgassing experiments.[12, 32, 33] Besides other techniques, photoelectron spectroscopy can potentially be used to track the EUV-induced chemical reactions in photoresists.

Photoelectron (or photoemission) spectroscopy (PES) is a powerful analytical technique that is capable of measuring both primary and secondary photoelectrons, in addition to the chemical composition of the material. In PES, an incident photon with energy greater than the work function of the material causes the emission of electrons. PES is in general a surface-sensitive technique as only the photoelectrons emitted within the first few nanometers from the surface can escape from the material and reach the detector. Indeed, on their way out of the material, the electrons lose kinetic energy due to scattering. The electron mean free path for electrons with kinetic energy ($E_K$) of 10–150 eV in solids is relatively short,[34, 35], being about 1–2 nm for a CAR,[36]. However, for lower energy electrons, $E_K < 5$ eV, the mean free path is larger making these electrons slightly sensitive to the bulk as well.[35]

PES can be performed using light sources in the X-ray region as well as EUV. Traditionally, X-ray photoemission spectroscopy (XPS) has been used to quantitatively study the chemistry of thin-film EUV photoresists.[36, 37] However, the high energy X-ray photons used in XPS can trigger additional chemical changes in EUV photoresists and the generated photoelectrons are not representative of the photoelectron cascade generated by EUV exposure.[38] Recently, PES using 13.5 nm EUV light has been deployed for measuring the generation of primary and secondary photoelectrons in EUV photoresist systems. By employing the exposure wavelength of 13.5 nm, key insights relevant to the EUV exposure process in lithographic scanners have been revealed. Several properties have been studied using EUV sources, such as efficiency of photoelectron production,[39, 40] the inelastic mean free path of primary photoelectrons,[36] as well as changes in photoelectron production vs exposure dose.[41, 42] These studies show that EUV photoemission yields information pertinent to the EUV lithographic exposure process. However, spectra obtained from EUV PES experiments are often complex to interpret. The photoelectrons generated from EUV exposure emerge from delocalized molecular orbitals, thus preventing a straightforward correlation between the photoemission peaks and the chemical composition of the resist. As an additional constraint, EUV photoemission at 13.5 nm has been so far limited to large-scale synchrotron facilities, thus limiting the accessibility of EUV photoemission for studying photoresist exposure dynamics.

In this work, we present a combined experimental and theoretical approach, which significantly advances EUV PES as a methodology for measuring chemical processes during EUV exposure in photoresist materials. EUV PES is performed on a variant of the environmentally stable chemically amplified photoresist (ESCAP) platform (a typical proxy for EUV CAR systems) using a coherent, tabletop EUV source based on high-harmonic generation (HHG) coupled with an advanced momentum microscope. The EUV exposure induces chemical changes in the photoresist, which result in modulation of the intensity of valence band peaks in the emission spectrum. These modulations are interpreted using an advanced simulation protocol that allows us to correlate changes in the EUV photoemission spectrum with chemical changes occurring in the model CAR system. This synergistic approach enables us to observe and track the degradation of a critical component responsible for the CAR exposure mechanism, the PAG anion ($PAG^-$). The $PAG^-$ is the precursor of the photoacid, which drives the acid-catalyzed deprotection mechanism that is responsible for the solubility switch, hence image formation, in CAR materials. The degradation of $PAG^-$ from EUV exposure could be one source of chemical stochastics underlying defect generation and ultimately reduction of device yield in EUV lithography processes. Our results not only pave the way for quantitative interpretation of 92 eV EUV exposure dynamics, but also open the door to time-resolved measurements of EUV exposure dynamics with coherent, femtosecond EUV sources.

## 2. METHODOLOGY

### 2.1 EXPERIMENTAL DETAILS

EUV PES at 13.5 nm excitation is performed using a coherent, femtosecond, table-top EUV system (XUUS 4, KM Labs) based on HHG, coupled with an advanced momentum microscope (KREIOS 150, SPECS) employing a hemispherical analyzer for photoelectron detection. The EUV source and photoemission measurement has been described in detail previously,[41, 43, 44]. Briefly, the photoelectron spectrum of the exposed ESCAP material was measured using the KREIOS 150 system with a pass energy of 100 eV and a slit width of 0.8 mm. Samples consisting of a modified version of the ESCAP material, described below, coated on a Si substrate were mounted on a metal plate with a conducting clip to minimize the effects of surface charging and transferred into the analysis chamber ($\sim 5 \times 10^{-10}$ mbar) of the KREIOS tool for measurements of the PES spectrum. The incident EUV beam power was attenuated to further reduce the effects of surface charging and to ensure that the flux is low enough to reduce the degradation of the photoresist during the first few spectra. The valence band PES spectrum was measured with an accumulated EUV dose of approximately 1–2 mJ/cm2.

X-ray photoelectron spectroscopy (XPS) measurements were performed on exposed ESCAP material at varying doses to track the relative loss of F-containing species in the film as a function of exposure dose. The XPS measurements were performed using a QUANTES XPS tool (Physical Electronics, PHI) employing a monochromatized Al-K$\alpha$ source (1486.6 eV) with a spot size of $\sim$100 µm rastered on a $500 \times 1000$ µm$^2$ to reduce the dose. Samples of the exposed ESCAP were cleaved to $1 \times 1$ cm$^2$ coupons and XPS measurements were made at the C 1s, N 1s, O 1s, S 1s, and F 1s absorption edges (which covers all atomic species in the film). An electron flood gun was employed for charge neutralization during the measurements to prevent charging-induced peak shifts from the insulating nature of the ESCAP material.

Residual gas mass spectrometry analysis was performed on the volatile products leaving the resist film during EUV exposure using an established EUV-induced outgassing tool, which has been described previously[31]. Briefly, the ESCAP material was spin-coated onto 200 mm wafers and EUV exposure was performed using a spectrally filtered plasma-discharge based z-pinch source (EQ-10, Energetiq). Mass spectra over the range of 0–100 amu were measured during EUV exposure using a Pfeiffer QMG422 quadrupole mass spectrometer, placed $\sim$5 cm from the exposed resist. The EUV exposure dose was limited to $\sim$7 mJ/cm$^2$ by raster scanning the EUV beam over the 200 mm wafer. Such low dose measurements help to track fragmentation and outgassing of volatile products at lithographically relevant doses. The same tool was also used to perform EUV exposures of the ESCAP for PES measurements. Samples exposed at 0, 40, and 200 mJ/cm$^2$ were also measured via EUV PES after the ellipsometry measurements were performed.

Photoresist samples were prepared via spincoating of the stock solutions. A modified version of the ESCAP was provided by FujiFilm with a composition as follows: a backbone copolymer (75.6 wt%) of *p*-hydroxystyrene (48 mol%) and *tert*-butyl methacrylate (52 mol%), (4- Methylphenyl) diphenyl sulfonium nonaflate as the PAG (17.3 wt%), and trioctylamine as the quencher (7.1 wt%). In all photoresist samples, the thickness of the photoresist was determined by fitting reflectance curves from a commercial spectroscopic ellipsometer (RC2, JA Woollam Company), which yielded thickness values of $\sim$30 nm for the ESCAP material, with a coating uniformity of ±0.5 nm. It is important to note that we do not use metallic underlayers as used in previous measurements of EUV PES on photoresist materials,[45, 46] as the additional electrons generated from the metallic underlayer can trigger further chemistries in the photoresist that would deviate from the expected behavior in a scanner environment.

### 2.2 COMPUTATIONAL DETAILS

Eight atomistic model structures of the unreacted ESCAP resist were generated using our in-house polymer builder python code based on coarse-grained molecular dynamics.[47] Each structure consists of about 1200 atoms. The simulation box was chosen as a cube of size 2.3 nm to match the measured density of 1.0 g/cm$^3$. The model structures were optimized using the Broyden–Fletcher–Goldfarb–Shanno (BFGS) algorithm[48] using forces calculated with density functional theory (DFT) using the PBEsol[49, 50] exchange correlation functional and imposing periodic boundary conditions as implemented in the CP2K software package.[51] To simulate the reacted resist, eight atomistic model structures were generated for each one of the five considered reaction mechanism depicted in Figure 1, i.e. acid formation from PAG; deprotection; deprotection with isobutene outgassing; deprotection with isobutene and CO$_2$ outgassing; and crosslinking. The reacted model structures were obtained from the optimized structures of the unreacted resist by replacing all reacting species with the corresponding products and removing the outgassing species. The reacted structures were then structurally reoptimized with the same protocol as for the unreacted structures.

To simulate the PES spectra, we rescaled the density of states (DOS) from DFT with the photoionization cross section. First, we computed Kohn–Sham energies with the HSE06[52] hybrid functional in combination with the standard DZVP basis set[53] and pseudo-potentials[54–56] provided in the CP2K software package.[51] The computed energies are given relative to the Fermi energy. To include the photoionization cross section in the simulated spectra, the following procedure was applied. Each Kohn–Sham orbital was first projected on individual atoms and angular momentum components to identify the atomic orbitals contributions, such as C2s, C2p, F2s, and so on. The contribution to the PES spectrum of each Kohn–Sham orbital is then computed as a Gaussian peak centered at the Kohn–Sham energy with an arbitrary width of 0.5 eV. The intensity of the Gaussian peak corresponding to each Kohn–Sham orbital is then computed by multiplying the occupation of the Kohn–Sham orbital by a weighted sum of all atomic orbital contributions, where the weights correspond to the coefficient of the atomic orbital decomposition multiplied by the photoionization cross section of the corresponding atomic orbital. The values of the photoionization cross section were obtained for each atomic orbital by a B-spline interpolation of the values tabulated in Ref. [57] for a photon energy of 92 eV. The interpolated values at 92 eV are reported in Table 1. The PES spectra are then obtained by merging the contributions of all Kohn–Sham orbitals. The PES spectra of the individual components of the ESCAP resist were obtained by merging the contributions of the Kohn–Sham orbitals of each molecular species. The PES spectra for the ESCAP and its components are the average over eight model structures.

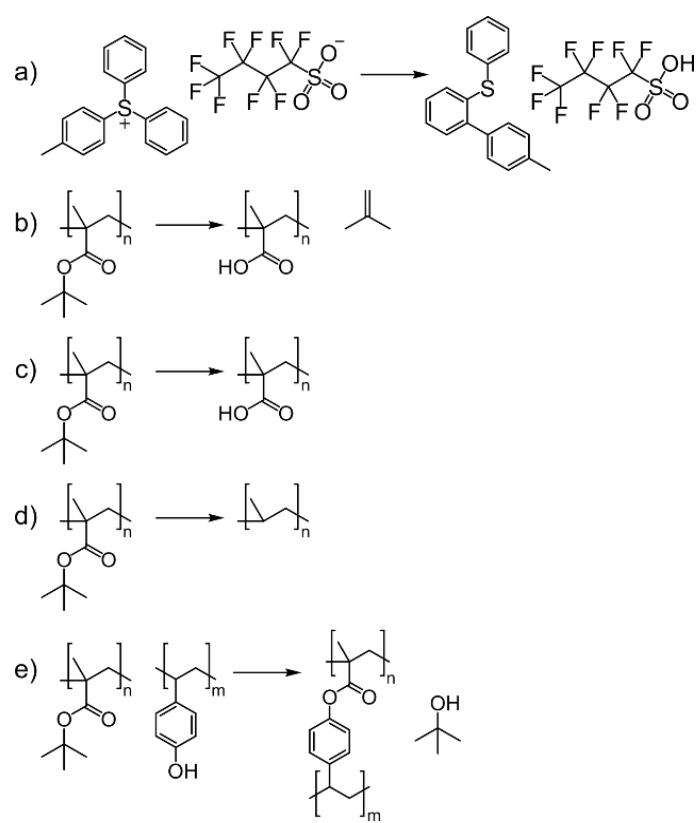

Figure 1. The five reaction mechanisms considered in this work: (a) acid formation from PAG; (b) deprotection of PBMA; (c) deprotection of PBMA with isobutene outgassing; (d) deprotection of PBMA followed by isobutene and $CO_2$ outgassing; and (e) crosslinking between PBMA and PHS. The schematic only shows the molecules left in the film, whereas the outgassed byproducts, which are removed from the simulation box, are not shown.

Table 1. Photoionization cross section for all the atomic elements of the CAR. Values are calculated for 92 eV photon energy by interpolating the values tabulated in Ref.[57].

| Orbital | Cross section ($10^6$ barn) |
|---------|------------------------------|
| H1s | 0.03 |
| C2s | 0.43 |
| C2p | 0.21 |
| N2s | 0.52 |
| N2p | 0.59 |
| O2s | 0.62 |
| O2p | 1.37 |
| F2s | 0.65 |
| F2p | 2.57 |
| S3s | 0.26 |
| S3p | 0.62 |

## 3. RESULTS AND DISCUSSION

The model ESCAP resist studied in this work consists of a random copolymer of two repeat units, namely *tert*-butyl methacrylate (PBMA) and *p*-hydroxystyrene (PHS), mixed with a photoacid generator (PAG) made of a cation-anion pair, a quencher molecule, and residues of the PGMEA and PGME solvents. The composition of the resist is shown in Figure 2(a). To investigate the chemical changes occurring during EUV exposure, the model resist was measured with PES before and after ex-situ exposure to EUV photons at two doses, 40 and 200 mJ/cm$^2$. The measured spectra are shown in Figure 2(b) and have also been published elsewhere.[41] Six main peaks can be distinguished in the experimental spectra of the unexposed resist. Here and in the following, we refer to the peaks numbered in decreasing energy from 1 to 6, i.e., 6 corresponds to the peak at the smallest binding energy (Figure 2(b)). Three of these peaks, namely peak 1, 4, and 5 undergo a visible decrease at high EUV dose (200 mJ/cm$^2$), whereas no significant variations are observed in the other regions of the spectrum.

The degradation of peak 1, 4, and 5 at a large exposure dose is an indication of EUV-induced chemical reactions occurring in the resist. Correlating the measured spectral changes with the underlying chemical mechanism is in general complicated due to the presence of overlapping contributions from different resist components. In this regard, first-principles simulations can provide guidance for the correct assignment of the spectral variations to the underlying EUV-induced reactions.

Figure 3 shows the comparison between the experimental and simulated spectra of the unexposed resist. The theoretical spectrum was computed as the average spectrum over 8 model structures with the nominal molar composition shown in Figure 2(a). For better comparison, the two spectra are aligned at the maximum by introducing a rigid horizontal shift, although the experimental and theoretical binding energies are, in principles, both referred to the Fermi edge. This shift can be attributed to a combination of factors, such as calibration errors in the experimental setup, charging effects, as well as the choice of basis set and functional in the computational protocol. It can be noticed that all six peaks observed experimentally are also present in the computed spectrum. Interestingly, the simulation reveals that peak 2 and 4 each result from the overlap of a pair of close-lying peaks, which cannot be experimentally resolved. In addition, it can be observed that the intensity of the measured spectrum increases steadily towards larger binding energies, contrary to the simulated spectrum (see Figure 3). This background effect in the experimental data is related to the loss of kinetic energy of the photoelectrons due to scattering mechanisms which are not included in the simulated spectra.

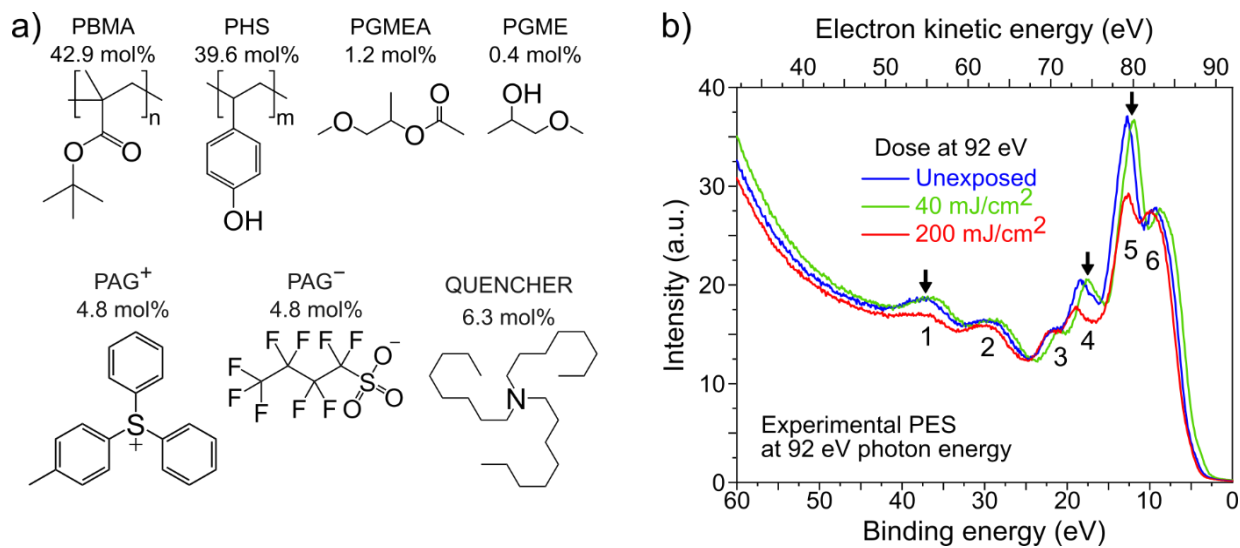

Figure 2. (a) Composition of the unexposed resist including residual solvents provided by Fujifilm. (b) Experimental photoelectron spectra for the unexposed resist and for the resist exposed to EUV at two exposure doses, 40 and 200 mJ/cm$^2$. Numbers 1–6 identify the most relevant peaks. Arrows highlight the degrading peaks 1, 4 and 5. The spectra have also been previously published in Ref. [41].

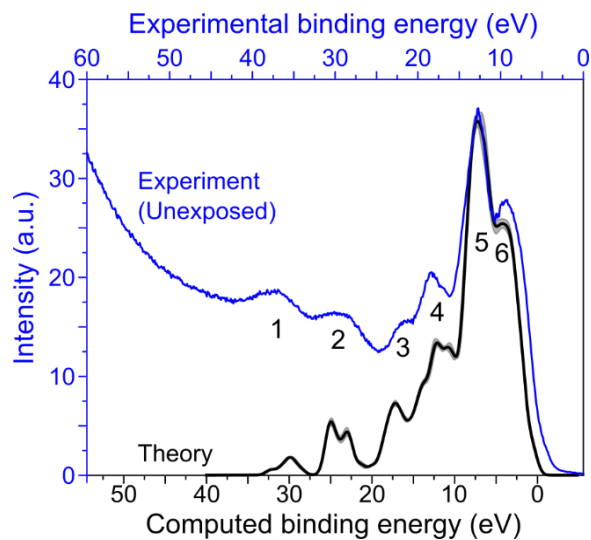

Figure 3. Computed spectrum compared with the experimental spectrum of the unexposed resist. Numbers 1–6 identify the most relevant peaks. The background in the experimental data is related to the loss of kinetic energy of the photoelectrons due to scattering mechanisms which are not included in the theoretical model.

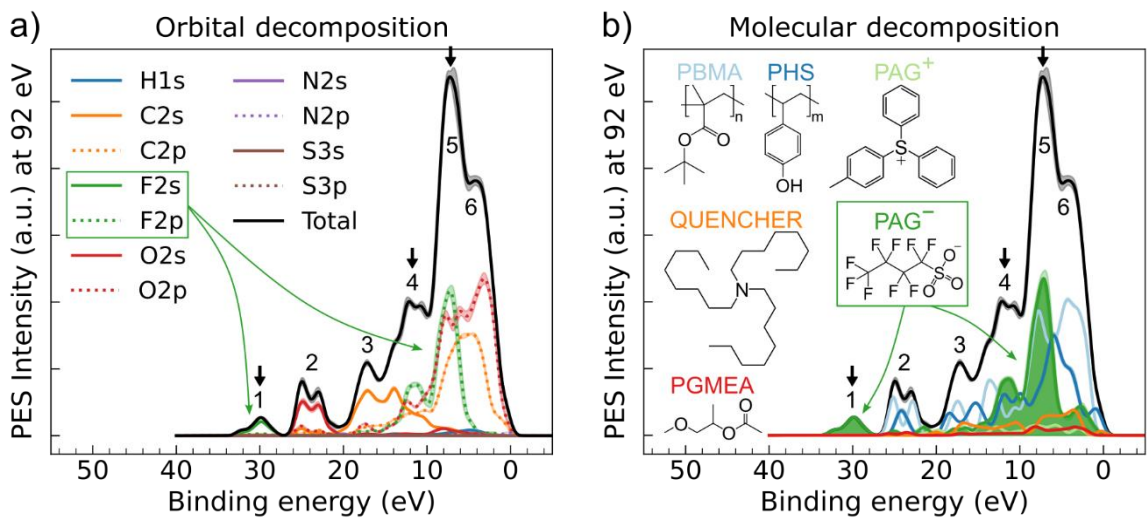

Figure 4. Computed spectrum of the unexposed resist highlighting the individual contributions of different (a) atomic orbitals and (b) molecular components. The spectra were computed assuming the nominal molar composition in Figure 2(a). Numbers 1–6 identify the most relevant peaks. Arrows highlight the degrading peaks 1, 4 and 5.

Figure 4(a) shows the decomposition of the spectrum in terms of the dominant atomic character of the molecular orbitals. The decomposition is obtained by projecting each molecular orbital onto atomic orbitals. From the decomposition, we can assign peak 1 ($\approx$ 30 eV) to F2s orbitals, peak 2 ($\approx$ 24 eV) to O2s and peak 3 ($\approx$ 17 eV) to C2s. All the other peaks result from a more complex interaction of different atomic orbitals, namely F2p, C2s, C2p and O2p for peak 4 ($\approx$ 11 eV), F2p, C2p and O2p for peak 5 ($\approx$ 7 eV) and C2p and O2p for peak 6 ($\approx$ 3 eV). Interestingly, the three peaks that decrease at high exposure doses, namely peak 1, 4, and 5, contain contributions from F2s and F2p. Conversely, the other peaks, that are not affected by exposure, do not contain contributions from fluorine orbitals. This correlation between peak degradation and fluorine-based orbitals (F2s and F2p) is a first indication that the reaction mechanism induced by EUV exposure is likely involving fluorine atoms.

To investigate the relationship between fluorine atoms and spectral changes, it is insightful to consider the contribution of each resist component to the PES spectrum, as shown in Figure 4(b). The decomposition is obtained by collecting the computed contributions of each molecular species in the model structure. Figure 4(b) shows that the degrading peaks 1, 4, and 5 contain contributions from the PAG$^-$ anion, whereas the non-degrading peaks mainly contain contributions from the other resist components. The correlation between the degrading peaks and the PAG$^-$ contribution is consistent with the correlation with F2s and F2p orbitals shown in Figure 4(a), as the PAG$^-$ anion is the only component of the resist containing fluorine atoms.

It is also interesting to note that the relatively large intensity of the PAG$^-$ contribution is due to the relatively large photoionization cross section of fluorine orbitals (see Table 1). Particularly, the cross section of F2p orbitals is one order of magnitude larger than that of C2p orbitals. This explains why the contributions of the fluorine-rich PAG$^-$ and of the carbon-rich PBMA in Figure 4(b) have comparable intensities, even though the molar ratio of PAG$^-$ is one order of magnitude smaller than that of PBMA.

The correlation between the degrading peaks 1, 4, and 5 and the PAG$^-$ contribution suggests that the PAG$^-$ anions undergo chemical changes when exposed to EUV photons. This is surprising, as PAG$^-$ molecules are expected to be chemically inert due to the high-stability of the C–F bonds. Indeed, PAG$^-$ anions belong to the family of polyfluoroalkyl substances (PFAS), which are popularly known as 'forever chemicals' due to their high resistance to degradation, which also makes them a hazard to human health and the environment.[58] To the best of our knowledge, the only reaction mechanism involving the PAG$^-$ that has been previously reported in the field of photolithography is the formation of perfluorobutanesulfonic acid by protonation.[20] However, the addition of a single proton to the PAG$^-$ anion is not sufficient to explain the significant peak decrease observed experimentally, as will be discussed more in details in Section 3.2. Instead, the measured spectral changes can be attributed to the removal of PAG$^-$ anions from the CAR system, as discussed in Section 3.1.

## 3.1 REMOVAL OF PAG⁻

Figure 5 shows the computed PES spectra of the CAR with increasing fractions of PAG⁻ removed. The spectra are computed as the sum of all the molecular contributions shown in Figure 4(b), after rescaling the intensity of the PAG⁻ contribution. Remarkably, the spectral changes for the release of PAG⁻ predicted from simulations correspond to the decrease of peak 1, 4, and 5, which are in qualitative agreement with the experimental peak decrease observed at increasing EUV doses. This result suggests that fluorine containing components, generated from the reaction of PAG⁻ molecules, leave the resist during EUV exposure. The degradation of PAG⁻ seems also confirmed by the decrease of the F1s peak measured with XPS, as shown in Figure 6(a).

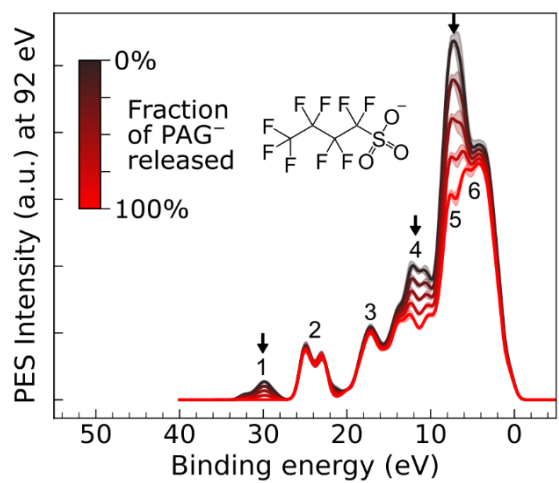

Figure 5. Computed spectra of the resist before and after removal of different fractions of PAG⁻ from the system. Numbers 1–6 identify the most relevant peaks. Arrows highlight the degrading peaks 1, 4 and 5.

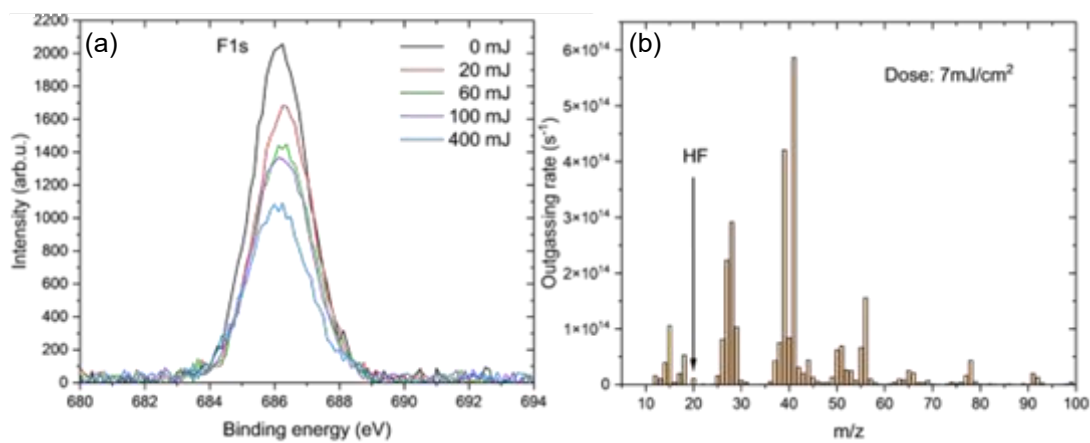

Figure 6. (a) F1s XPS of the unexposed resist and of the resist exposed ex-situ to EUV at four increasing doses: 20, 60, 100 and 400 mJ. The spectrum shows a decrease of the F1s peak, which can be attributed to the EUV-induced breakdown and removal of fluorine-rich PAG⁻ molecules from the sample. (b) RGA mass spectrum of the molecular fragments outgassed from the resist during exposure to EUV radiation at a low dose of 7 mJ/cm$^2$. The signal at 20 m/z can be attributed to HF molecules which are probably formed from the breakdown of PAG⁻.

The details of the PAG⁻ degradation reaction are not fully understood. A possible mechanism is the EUV-induced cleavage of the C–S, C–C, and C–F bonds in the PAG⁻ molecules, leading to the formation of smaller molecules such as $SO_3$, HF, and shorter polyfluoroalkyl compounds, which may leave the resist in gaseous form. The release of HF molecules has also been detected by complementary residual gas analysis (RGA) on the EUV-exposed CAR (see Figure 6(b)).[12] The formation of $SO_3$, HF and polyfluoroalkyl compounds compounds has also been reported for exposure to different types of ionizing radiation, such as γ-rays or electron beam.[59–62]

In view of photolithographic applications, the breakdown and outgassing of PAG⁻ anions is expected to be detrimental. The PAG⁻ anion is the precursor of the acid species which enables the amplification of the solubility switch mechanism through the deprotection of acid-labile protecting groups on PBMA copolymer units. Therefore, the PAG⁻ degradation would likely result into a less efficient deprotection reaction, hindering the patterning quality. Yet, since PES is a surface-sensitive method, it is not clear whether the removal of PAG⁻ only affects the surface or also the bulk.

### 3.2 ADDITIONAL REACTIONS

In this section we investigate whether other reaction mechanisms, besides the breakdown of PAG⁻ molecules, may explain the observed degradation of peaks 1, 4, and 5 upon exposure to EUV photons. To unravel the mechanisms induced by EUV photons, we use simulations to compute the spectra using a bottom-up approach: first, a set of possible reaction mechanisms is selected, based on an educated guess from the literature; then, atomistic model structures of the reacted resist are generated, for each considered reaction, by replacing all reactants with the products; finally, the spectra of the reacted models are computed and compared with the experiment. A schematic workflow of the simulation protocol is depicted in Figure 7.

Based on the discussion above and Figure 4(b), it is clear that the degrading peaks 1, 4, and 5 in the experimental spectra are mostly due to the contribution of PAG⁻ and the copolymer. Therefore, it is expected that the underlying reaction mechanisms involve either or both these resist components. We thus consider three reaction mechanisms involving either the copolymer or the PAG, namely (i) the reaction of PAG leading to the formation of perfluorobutanesulfonic acid (Figure 8(a)), (ii) the deprotection reaction of PBMA, followed by outgassing of isobutene and $CO_2$ (Figure 8(b)), and (iii) the crosslinking reaction between PBMA and PHS copolymer units (Figure 8(c)).

The first considered reaction mechanism (Figure 8(a)) is the formation of the acid from the PAG. This reaction can occur either from a neutral excitation of the cation induced by energy transfer from a high-energy secondary electron or through dissociative attachment of a low-energy secondary electron followed by electron release.[20] First, the cation rearranges into neutral phenylthiobiphenyl through the cleavage of a C–S bond, followed by the deprotonation of a phenyl group. Then, the proton is transferred to the anion to form perfluorobutanesulfonic acid. This acid is believed to facilitate the deprotection reaction of the PBMA copolymer units, resulting in the amplification of the solubility switch in CAR materials.[8, 20, 63]

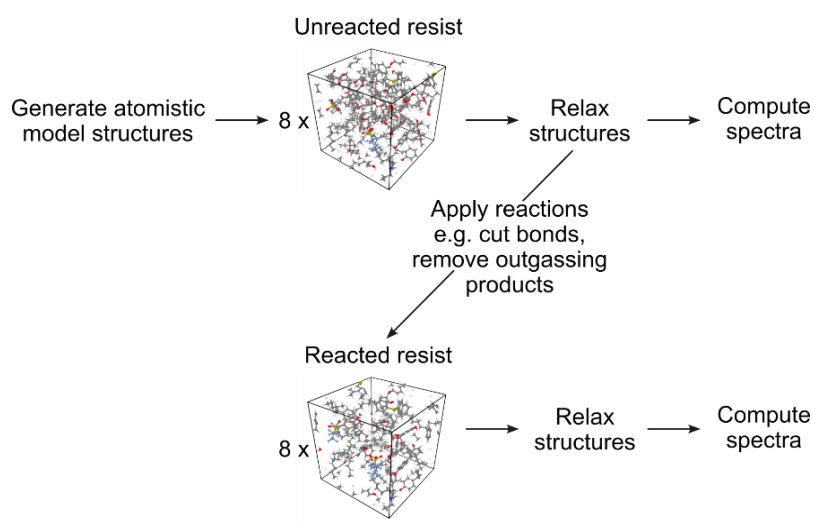

Figure 7. Schematic workflow of the simulation protocol to compute the PES spectra of the resist before exposure (unreacted) and after exposure (reacted).

As shown in Figure 8(a), the computed PES spectra of the CAR undergo only small changes after the formation of acid molecules from the PAG. Indeed, most peaks remain unchanged, except for peak 4, which becomes narrower and more intense, and peak 5, which gets narrower and shifts to the left. None of these changes seem comparable with the significant decrease of peaks 1, 4, and 5 measured experimentally. Moreover, the predicted changes for the reaction of PAG are smaller than the experimental resolution. Therefore, the occurrence of acid formation, which is likely to happen upon exposure, can be neither confirmed nor excluded from experimental PES spectra alone.

The second reaction mechanism (Figure 8(b)) is the deprotection of PBMA, followed by outgassing.[8, 12, 31, 63–67] We considered three reaction steps and generated three corresponding sets of model structures, of which we computed the PES spectra. The first step is the deprotection reaction of PBMA, which is the main mechanism leading to the solubility switch of the resist. This step, which is acid catalyzed, consists in the cleavage of the apolar tert-butyl protecting group, which is an ester elimination reaction, from the PBMA unit, resulting in a polar carboxyl group on the copolymer chain. The tert-butyl cation is acidic and tends to transform into isobutene releasing a proton that then catalyzes the next deprotection reaction. The second step is the removal of isobutene from the system. Indeed, outgassed isobutene cations have been detected in RGA and photoelectron photoion coincidence (PEPICO) experiments.[12, 14, 33, 67] Finally, the third step consists in the Norrish type 1 scission of the PBMA side-chain through a C–C bond cleavage accompanied with a proton transfer from the carboxyl oxygen to the backbone. The results of this final step are the bare polymer backbone and a $CO_2$ molecule. This reaction has been confirmed for EUV-exposed CAR by RGA and PEPICO experiments.[12, 14, 33, 67] For the sake of simplicity, we consider the limiting case of complete outgassing and therefore remove all $CO_2$ molecules from the reacted model structures. We notice that the reaction leading to the $CO_2$ release is likely detrimental for the resist performance as it leads to the unwanted solubility switch of the PBMA copolymer unit from the polar carboxyl group to the apolar alkyl backbone, which would make the resist insoluble to the developer.

By looking at the computed spectra in Figure 8(b), it appears that the measured spectral changes upon exposure cannot be explained by the degradation of PBMA alone. Indeed, none of the three considered reaction steps, namely deprotection, outgassing of isobutene, and complete side-chain scission with $CO_2$ release, lead to the simultaneous decrease of peaks 1, 4, and 5 as observed experimentally. On the contrary, the first two steps, i.e. deprotection and isobutene outgassing, do not induce visible alterations of the computed spectra. Only the third step, which is the complete side-chain scission of PBMA accompanied with $CO_2$ outgassing, induces large modifications in the computed spectra, leading to an overall decrease of peaks 2, 3, 4, 5, and 6, but not 1.

The third reaction mechanism (Figure 8(c)) considered in this study is the crosslinking of PBMA and PHS copolymer units. The reaction involves the cleavage of the hydroxyl group of PHS and of the tert-butyl protecting group from the PBMA unit, leading to the formation of tert-butyl alcohol and a crosslinked species which was previously proposed by Blau et al.[68] The formation of crosslinking bonds in the copolymer matrix has been suggested as the responsible mechanism for the unwanted solubility switch observed at high EUV doses, although no direct evidence of crosslinking has been reported to the best of our knowledge.

Interestingly, the computed spectra for the crosslinked CAR in Figure 8(c) show that only limited changes can be observed in the spectra upon crosslinking. The changes mostly involve peak 2, 3, and 5, which appear to broaden towards larger binding energies. The predicted spectral variations are small and beyond the experimental resolution. Therefore, crosslinked products cannot be confirmed nor excluded through PES spectra, although it is clear that crosslinking reactions alone are not sufficient to explain the observed decrease of peak 1, 4, and 5.

Overall, none of the reactions displayed in Figure 8 induces theoretical spectral changes that are comparable with the decrease of peaks 1, 4, and 5 observed experimentally (Figure 2(b)). The peak degradation upon EUV exposure is therefore attributed to the loss of $PAG^-$ from the resist, as discussed in Section 3.1. The acid formation, deprotection and crosslinking reactions cannot be detected directly from PES, therefore complementary experimental techniques are required to investigate these reactions.

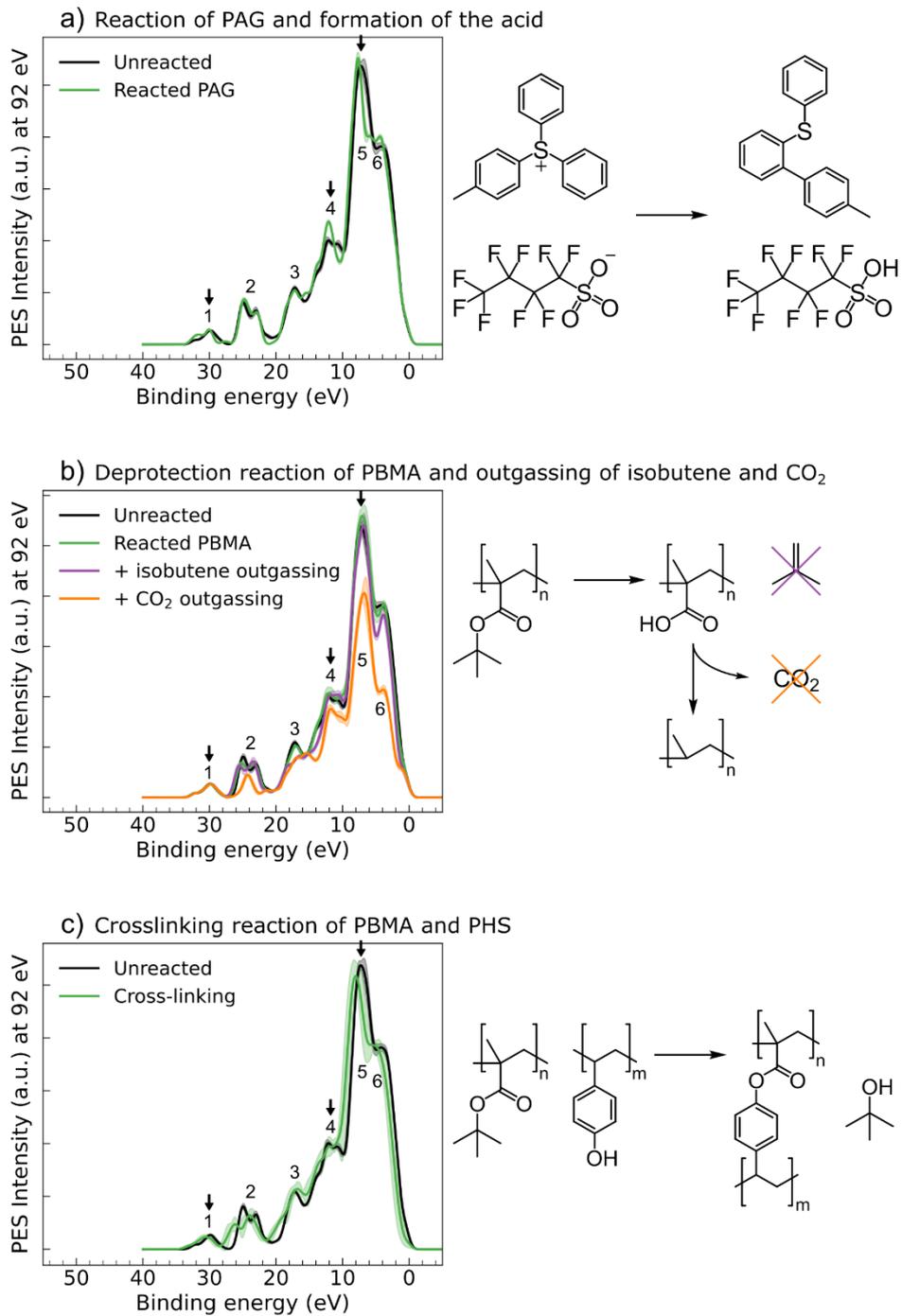

Figure 8. Computed spectra of the resist before and after three reactions: (a) degradation of all PAG molecules leading to acid formation, (b) deprotection of all PBMA monomers accompanied by outgassing of isobutene and $CO_2$ byproducts, and (c) crosslinking reaction between the PBMA and PHS monomers. Numbers 1–6 identify the most relevant peaks. Arrows highlight the degrading peaks 1, 4 and 5.

## 4. CONCLUSIONS

In this work, PES spectra were measured on a model CAR using 92 eV photons, before and after ex-situ exposure to increasing EUV doses, to investigate the chemical changes occurring in the resist during EUV photolithography. The spectra show a clear decrease of three distinct peaks at high doses. To interpret these spectral changes, we computed the theoretical PES spectra of the resist using first-principles simulations. By comparing computed and experimental spectra, we could attribute the peak decrease to chemical changes in the PAG$^-$, a fluorine-rich constituent of the resist and the precursor of the photoacid, which is crucial for catalyzing the deprotection reaction leading to the solubility switch mechanism that is needed for pattern formation in photolithography. More specifically, the spectral changes can be attributed to a reaction mechanism that has to the best of our knowledge, not been previously observed; namely, the breakdown and release of PAG$^-$ molecules from the surface of the resist materials. Although the details of the PAG$^-$ degradation are not fully understood, it is possible that the molecule breaks into smaller fluorine-rich fragments that then escape the material. One such fragment being HF, was detected in complementary RGA experiments. The removal of PAG$^-$ from the resist is also consistent with the decrease of the F1s peak recorded at increasing exposure doses in complementary XPS measurements. To strengthen the interpretation of the PES spectra and rule out other possible explanations for the measured spectral changes, we have also computed the theoretical spectra of the resist for three other possible reaction pathways, namely, the generation of photoacid, the deprotection of the copolymer followed by outgassing, and crosslinking. It was found that none of these reactions explain the measured spectra, confirming that the spectral changes can be attributed to the PAG$^-$ breakdown.

Overall, the results of this work provide a comprehensive overview of the reaction mechanisms in CARs upon EUV exposure and reveal the degradation and removal of the PAG$^-$ anion from the resist film, which may be a potential source of stochastic defects that has not been previously considered. Moreover, this work shows that the combination of experimental PES with first-principles calculations can provide a unique platform for unraveling the complex exposure mechanisms in EUV photoresists.

## ACKNOWLEDGEMENTS

D. P. S. and K. M. D. acknowledge funding from the European Union's Horizon 2020 research and innovation program under the Marie Sklodowska-Curie grant agreement No.'s 101032241 (D.P.S.) and 101031245 (K.M.D.). We also gratefully acknowledge FujiFilm for providing the model ESCAP material used in this work. We would also like to thank Esben W. Larsen, Roberto Fallica, and Danilo De Simone for their support and assistance.